\documentclass[Crown,sageh,times]{sagej}
\usepackage{moreverb,url}

\usepackage[colorlinks,bookmarksopen,bookmarksnumbered,citecolor=red,urlcolor=red]{hyperref}
\usepackage{subfigure}
\usepackage{physics}
\usepackage{threeparttable}
\newtheorem{prob}{Problem}
\newtheorem{theo}{Theorem}
\newtheorem{ex}{Example}
\newtheorem{theorem}{Theorem}

\newtheorem{lemma}[theorem]{Lemma}
\newtheorem{assum}{Assumption}
\newtheorem{definition}{Definition}
\newtheorem{remark}{Remark}

\newcommand\BibTeX{{\rmfamily B\kern-.05em \textsc{i\kern-.025em b}\kern-.08em
T\kern-.1667em\lower.7ex\hbox{E}\kern-.125emX}}

\def\volumeyear{2016}

\begin{document}

\runninghead{Bhargavi Chaudhary \textit{et~al.}}

\title{An Event-based State Estimation Approach for Positive Systems with Positive Observers}

\author{Bhargavi Chaudhary\affilnum{1}, Krishanu Nath\affilnum{2}, Subashish Datta\affilnum{1}, and Indra Narayan Kar\affilnum{1}}

\affiliation{\affilnum{1}Department of Electrical Engineering, Indian Institute of Technology Delhi, India, 110016.\\
\affilnum{2}Department of Electrical Engineering,
Dr B R Ambedkar National Institute of Technology Jalandhar, India.}

\corrauth{Bhargavi Chaudhary, Department of Electrical Engineering, Indian Institute of Technology Delhi, Hauz Khas-110016, New Delhi, India.}

\email{eez208419@ee.iitd.ac.in}

\begin{abstract}
This article addresses the problem of state observer design for continuous-time linear positive networked systems. Considering the bandwidth constraint in the communication network, an event-measurement-based positive observer design is proposed. The physical interpretation of a positive observer differs from that of a general observer. Its primary goal is to ensure that all state estimates remain non-negative at all times.
Using output measurements, a law with weighted sampling error is used to determine the sampling sequence between the system and the observer. The observer dynamics are designed using the standard Luenberger structure with the event-based sampled output information, which is updated only when an event occurs. Assuming observability and sufficient conditions for the positivity of the system, the asymptotic stability of the observer dynamics with sampled information is established. Sufficient conditions of stability and positivity are derived using linear matrix inequalities. Moreover, the design ensures that the event-based architecture is free from Zeno behavior, ensuring a positive minimum bound on the inter-execution time. In addition, numerical simulations on a three-tank system having variable cross-sections are used to demonstrate the efficacy of the proposed event-based positive observer.
\end{abstract}


\maketitle

\section{Introduction}
\label{sec:introduction}
Positive systems are certain categories of systems where the system states and the outputs consistently have non-negative values, given that the initial conditions and inputs are also non-negative \cite{farina2000}. Due to their numerous applications in a variety of domains, such as network communication \cite{Shorten2006}, biological engineering \cite{haddad2005stability}, and population studies \cite{Caswell2001}, these systems have drawn a lot of attention. Researchers have established important criteria for analyzing positive systems, such as controllability, observability, and realization \cite{FARINA19951,Valcher20091586}, which provide insights into how these systems can be influenced and monitored.

In control engineering, a fundamental challenge lies in accurately determining the states of a system. Often, directly measuring all these states is impractical or even impossible. To address this, state observer design plays a vital role. The Luenberger observer stands as a benchmark solution, providing a way to estimate the true state of a system \cite{Luenberger1971596}. Research on observer design for positive systems has focused on maintaining positivity in the observer itself. Pioneering works \cite{van1998positive,hardin2007observers,back2008design} laid the foundation for this concept. Advancements include positive observers for delayed systems \cite{zaidi2014static}, bounded-state observers for full state tracking \cite{rami2007controller}, and even those handling missing output data \cite{WANG2015427}.

Modern control systems leverage communication networks to connect geographically separated components. While this offers flexibility, it introduces challenges like network delays and limited bandwidth \cite{Garcia2013,Batmani2021}. These limitations can significantly impact control performance. In the recent developments, some of these challenges are dealt by exploring event-triggered control schemes \cite{PENG2018113, Cucuzzella2020,wangieeesensor}. This approach reduces communication overhead by transmitting data only when specific pre-defined events occur, determined by the sensor based on current measurements. This strategy can significantly improve energy efficiency in networked control systems.

These systems experience a network burden or dropped data, making observer design a new and interesting challenge. The landscape of control systems has evolved with the introduction of networked control systems. These systems rely on communication channels to transmit data between sensors and controllers. As a consequence, minimizing the amount of data transmitted becomes a critical objective. Some existing research focuses on this challenge within the realm of networked control systems by exploring event-based observers \cite{Song2020} where the goal is to design efficient methods for state estimation while minimizing communication traffic. Few studies that delve into various categories of networked control systems and have proposed tailored solutions for each type to achieve optimal state estimation with minimal network burden can be found in \cite{Batmani02122021, ZHANG20141852,song2018}. These articles investigated observer design for general linear systems without considering positivity constraints on state variables. A key gap remains in designing positive observers specifically for positive networked control systems. It is important to note that the concept of a positive observer differs significantly from that of a general observer. The key objective of a positive observer is to ensure that both the state estimates and the estimation error remain non-negative at all times. This characteristic has recently drawn increased attention from researchers.
However, positive observer design on event-based measurement for continuous-time networked positive systems has not been discussed yet, motivating the current research on this gap. 

In this work, an event-based positive state observer is designed for a positive linear system based on network communication. The main contributions of the proposed design approach are enumerated as follows:
\begin{itemize}
    \item [(i)] A novel positive observer with event-based measurement law is proposed for a networked positive linear system.
    \item [(ii)] For ensuring the positivity and asymptotic stability of the observer with sampled information, sufficient conditions are established in the Linear Matrix Inequality (LMI) framework.
    \item [(iii)] A positive minimum bound on Inter-Execution Time (IET) is ensured through the proposed design in order to exclude Zeno behaviour.
    \item [(iv)] The proposed observer is verified through simulation on a three-tank system having variable cross-section.

\end{itemize}


\textit{Notation:}
In the manuscript, the following notations are used. $\mathbb{R}^n$ and  $\mathbb{R}_+^n$ stand for $n-$dimensional vector defined over the set of real numbers and positive real numbers, respectively. $\mathbb{R}^{m\times n}$ and $\mathbb{R}_+^{m\times n}$ represent matrices of dimension $m\times n$ over set of real number and positive real number. $A\geq 0$, $A>0$, $A\leq 0$, and $A<0$ denote that the elements of matrix $A$ are all non-negative, positive, non-positive, and negative, respectively. Similarly, for vector $a\geq b$ denotes that the elements $a_i$ of vector $a$ are all greater than or equal to the elements $b_i$ of vector $b$. $A\succ 0$ and $A\prec 0$ denote the positive and negative definiteness of a matrix, respectively. $A^\top$ denotes the transpose of a matrix $A$. $\|A\|$ represents Euclidean norm of a matrix $A$. $\ast$ represents a symmetric matrix.

\section{Preliminaries and Problem formulation}
Consider a continuous time linear system as:
\begin{align} \label{PS_eq}
\begin{split}
     \Dot{x}(t)&=Ax(t); \quad x(0)=x_0,\\
        y(t)&=Cx(t),
        \end{split}
\end{align}
where system state $x(t) \in \mathbb{R}^n$ 
and output $y(t) \in \mathbb{R}^r$. Matrices $A$ and $C$ are respectively the state and output matrices of the system with proper dimensions. The positivity of the system is defined as follows. 
\begin{definition}
       For any initial condition $x_0 \geq 0$, 
       the system \eqref{PS_eq} is defined as positive system if its corresponding trajectory $x(t) \geq 0$ and $y(t) \geq 0$ for all $t \geq 0$. 
\end{definition}
For the continuous-time positive system \eqref{PS_eq}, this work aims to design an event-based positive observer to estimate the state $x(t)$. Before formally defining the problem, we first present key definitions and results on positive systems that will be utilized throughout the paper.
\begin{definition}
    A matrix $M \triangleq[m_{ij}] \in \mathbb{R}^{n \times n}$ is defined as a Metzler matrix if its off-diagonal elements are non-negative i.e., $m_{ij} \geq 0$, $i \neq j$.
\end{definition}
The checkable condition for the positivity of the system is provided by a classical result \cite{luenberger1979introduction} as in the following Lemma.
\begin{lemma} 
    The system \eqref{PS_eq} is positive if and only if system matrix $A$ is Metzler and $C\geq 0$.
\end{lemma}

\begin{lemma} \cite{farina2000} \label{lemma_metzler}
    For a Metzler matrix $A$, there always exists a sufficiently large constant $\lambda>0$ such that $A+
\lambda I\geq 0$.
\end{lemma}
The result of the asymptotic stability analysis of the positive system \eqref{PS_eq} is presented as follows.
\begin{lemma} \cite{rami2007controller} \label{Hurwitz_stable}
    Considering a continuous time positive linear system \eqref{PS_eq}, the following assertions are equivalent:
    \begin{enumerate}
    \item System matrix $A$ is a Hurwitz matrix.
    \item System is asymptotically stable for every non-negative initial conditions.
    \item System is asymptotically stable if and only if there exists a diagonal positive definite matrix $D$
    such that $A^\top D + DA \prec 0$.
    \item There exists a positive $\lambda$ such that $A^\top\lambda<0$.
\end{enumerate}
\end{lemma}

\begin{assum}
The pair $(A,C)$ is observable and positive system \eqref{PS_eq} is asymptotically stable.
\end{assum}

\begin{figure}[h]
\centering
\includegraphics[width=0.4\textwidth]{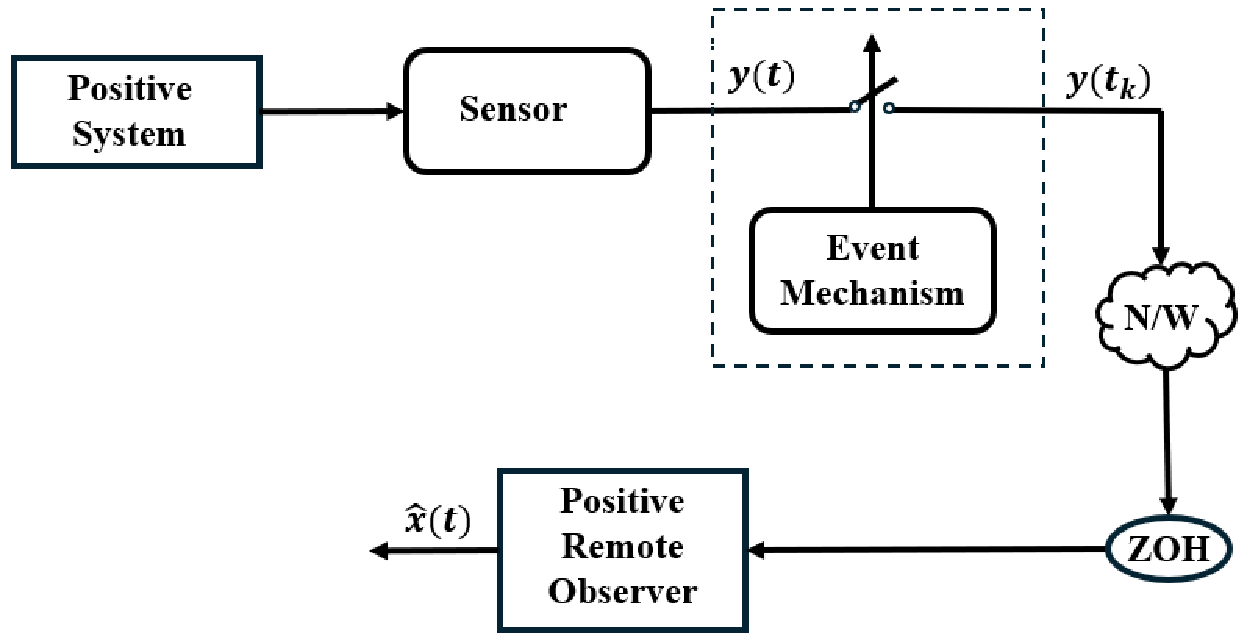}
\caption{Schematic of the network-based positive state observer for a positive system \eqref{PS_eq}}
\label{ETM_scheme}
\end{figure} 

\section{MAIN RESULTS}
Consider the block diagram shown in Figure \ref{ETM_scheme}, where the output $y(t)$ of the positive system \eqref{PS_eq} is continuously available for measurement. The output of the system is transmitted to a positive observer, where the system states need to be reconstructed. The conventional time-based sampling approach often leads to frequent instances of transmission of the sampled output $y(t)$. To reduce the number of transmissions in the network, sampling and information exchange between the system and the observer must be done using a need-based approach. This can be achieved by using the sampling method where the decision of sampling is based on the evolution of the output of the system \cite{kawaguchi2017state}. 

The event generator monitors the weighted output sampling error which is the difference between the scaled previous output, $\beta y(t_k)$ and the current output, $y(t)$ as given by:
\begin{align}\label{eq:sampling error}
    \epsilon(t)=\beta y(t_k)-y(t)\geq 0, \quad t\in [t_k,t_{k+1})
\end{align}
 where
$\beta>1$ is a design scalar to ensure $\epsilon(t)\geq 0$ and $y(t_k)$ is the sampled output at the last sampled instant $t_k$. It is always possible to find a sufficiently large $\beta$ such that the weighted output sample error $\epsilon(t)\geq 0$. Here $t_k$ with $k \in \mathbb{N}_0$ are sampling instances with the first sampling $t_0=0$.
 
We assume that $\epsilon(t)$ for all $t\in[t_k,t_{k+1})$ satisfies the following condition:
\begin{equation} \label{eq:event_law}
\begin{aligned}
    \epsilon(t)&\leq \alpha\beta y(t)+\beta y(t)-y(t)\\
    \epsilon(t)&\leq (\alpha\beta+\beta-1)y(t)
    \end{aligned}
\end{equation}
where $\alpha>0$ is a threshold parameter.

The event generator will transmit the data packet $y(t_k)$ only if the specified condition in \eqref{eq:event_law} is violated, i.e., if for any $i\in \mathcal{N}$ the following condition holds,
\begin{equation} \label{eq:event_violation_law}
    \epsilon_i(t)\geq (\alpha\beta+\beta-1)y_i(t)
\end{equation}
then we flag the time instant as $t_{k+1}$ for transmitting the data next to the $k^{th}$ instant, which is defined as follows:
\begin{equation}\label{eq: triggering rule}
    t_{k+1}=\inf \{t>t_k|~\epsilon_i(t)>(\alpha\beta+\beta-1)y_i(t)\}.
 \end{equation}

\begin{remark}
    The event law \eqref{eq:event_law}, motivated by \cite{ETM_brief}, ensures that the next sampling occurs as soon as the error grows beyond the predefined threshold value $(\alpha\beta+\beta-1)y(t)$.
    This guarantees that the sampling error is always bounded for an interval $t\in [t_k,t_{k+1})$.
\end{remark}

The event-based positive observer for the positive system \eqref{PS_eq} with the sampled information of the output for $t\in[t_k,t_{k+1})$
is designed as follows:
\begin{equation}  \begin{aligned} \label{obs}
\dot{\hat{x}}(t)&= A \hat{x}(t)+\beta L y(t_k)-L\hat{y}(t),
 \\ \hat{y}(t)&=C \hat{x}(t),
\end{aligned}
\end{equation}
where $\hat{x}(t)$ is the estimate of state $x(t)$, $\hat{y}(t)$ is the observer output and $y(t_k)$ is the data packet received by the observer. $L \in \mathbb{R}^{n \times r}$ is the observer gain matrix to be designed. 

Putting  $\beta y(t_k)=\epsilon(t)+y(t)$ from \eqref{eq:sampling error} in \eqref{obs}, we have
\begin{equation}
\dot{\hat{x}}(t)=( A-LC) \hat{x}(t)+L\epsilon(t)+ L y(t).    
\end{equation}
Let the estimation error be defined as $e(t)=\hat{x}(t)-x(t)$, then the error dynamics of the observer are described as follows:
\begin{equation}  \begin{aligned}
 \dot{e}(t)&=( A-LC)e(t)+L\epsilon(t).
 \end{aligned}
\end{equation}
Now defining $\xi(t)\triangleq\begin{bmatrix}
    x^\top(t) & e^\top(t)
\end{bmatrix}^\top $, the augmented system can be described as:
\begin{equation} \label{Closed-loop_sys}
\dot{\xi}(t)=\begin{bmatrix}
    A &0\\0 & A-LC
\end{bmatrix} \xi(t)+\begin{bmatrix}
    0\\L
\end{bmatrix}\epsilon(t).  
\end{equation}
The main objective of this paper is to design an event-based positive observer such that observation error, $e(t)=\hat{x}(t)-x(t)$, converges asymptotically to zero, i.e., $\lim_{t\rightarrow 0} e(t)=0$. We summarize the requirement as follows:

\begin{prob} \label{problem}
    For a positive system \eqref{PS_eq}, design an event-based positive observer gain matrix $L$ with an event law \eqref{eq:event_law} such that the augmented system in \eqref{Closed-loop_sys} is positive and asymptotically stable. Also, the system does not exhibit Zeno behavior.
\end{prob}

The sufficient conditions for the positivity and asymptotic stability of the system \eqref{Closed-loop_sys} are given in Theorem \ref{theorem_1}, which leads to the design of a positive observer for the given positive system. Taking into consideration this criterion, the positive state observer design is presented. Subsequently, for the non-existence of Zeno behavior, a sufficient condition is proposed in Theorem \ref{theorem_2}.
\begin{theo} \label{theorem_1}
 Problem \ref{problem} has a solution if for constants $\alpha>0$ and $\beta>1$, there exist a real constant $\lambda>0$, positive definite diagonal matrices $P$, $Q$ and  a matrix $W>0$, such that the following holds,
\begin{align}
   \begin{bmatrix}
    PA+A^\top P &(\alpha\beta+\beta-1)C^\top W^\top\\ \ast
    &QA+A^\top Q-WC-C^\top W^\top
\end{bmatrix} &\prec 0 \\
 QA-WC+\lambda QI&\geq0.
 \end{align}   

Under these conditions, the positive observer gain matrix $L$ is given as $L=Q^{-1}W$.
\end{theo}
 \proof
The closed-loop system in \eqref{Closed-loop_sys} can be rewritten as follows:
\begin{equation} \label{Closed-loop_sys_1}
\dot{\xi}(t)=\bar{A}\xi(t)+\bar{L}\epsilon(t), \end{equation}
where $\bar{A}=\begin{bmatrix}
    A &0\\0 & A-LC
\end{bmatrix}$ and $\bar{L}=\begin{bmatrix}
    0\\L
\end{bmatrix}$.
\\ \\
The stability analysis of system \eqref{Closed-loop_sys} is given below.\\
Consider the Lyapunov candidate as:
    \begin{equation}
    \label{eq:Lyapunov_function}
V=\xi^\top(t)D\xi(t).
    \end{equation}
    where $D$ is a diagonal positive definite matrix.
Taking the time derivative of \eqref{eq:Lyapunov_function} along the trajectories of \eqref{Closed-loop_sys_1}
 \begin{equation}
     \dot{V}=\xi^\top(t) D \dot{\xi}(t)+\dot{\xi}^\top(t)D\xi(t).
 \end{equation}   
From \eqref{Closed-loop_sys_1},
\begin{align}
     \dot{V}&=\xi^\top(t) D (\bar{A}\xi(t)+\bar{L}\epsilon(t))+(\bar{A}\xi(t)+\bar{L}\epsilon(t))^\top  D\xi(t), \nonumber \\
 &=\xi^\top(t) (D\bar{A}+\bar{A}^\top D)\xi(t) +\xi^\top(t) D \bar{L}\epsilon(t)+\epsilon^\top(t)\bar{L}^\top D\xi(t), \nonumber \\
 &=\xi^\top(t) (D\bar{A}+\bar{A}^\top D)\xi(t) +2\xi^\top(t) D \bar{L}\epsilon(t).
\end{align}
From the event law \eqref{eq:event_law}, the above equality can be reduced to
 \begin{align}
  \dot{V}&\leq\xi^\top(t) (D\bar{A}+\bar{A}^\top D)\xi(t) +2\xi^\top(t) D \bar{L}(\alpha\beta+\beta-1)Cx(t).
 \label{Vdot_term}
\end{align}
Let $D=\begin{bmatrix}
        P & 0 \\
        0 & Q 
    \end{bmatrix}$ where $P$ and $Q$ are diagonal positive definite matrices, then $D\bar{L}=\begin{bmatrix}
        P & 0 \\
        0 & Q 
    \end{bmatrix} \begin{bmatrix}
        0 \\
        L
\end{bmatrix}=\begin{bmatrix}
        0 \\
        QL 
\end{bmatrix}.$\\ \\
The 1st term $\xi^\top(t) (D\bar{A}+\bar{A}^\top D)\xi(t)$ in \eqref{Vdot_term}  can be represented as,
\begin{equation} \label{eq:1st_term}
    \xi^\top(t)\begin{bmatrix}
    PA+A^\top P &0\\ \ast &QA+A^\top Q-QLC-C^\top L^\top Q
\end{bmatrix}\xi(t).
\end{equation}
The 2nd term $2\xi^\top(t) D \bar{L}(\alpha\beta+\beta-1)Cx(t)$  in \eqref{Vdot_term}  can be represented as,
\begin{equation} \label{eq:2nd_term}
    \xi^\top(t)\begin{bmatrix}
        0 &(\alpha\beta+\beta-1) C^\top L^\top Q \\
       \ast & 0
    \end{bmatrix}\xi(t).
\end{equation}
Combining both \eqref{eq:1st_term} and \eqref{eq:2nd_term} we have
\begin{equation}
    \dot{V}=\xi^\top (t) \left[\begin{smallmatrix}
       PA+A^\top P  &(\alpha\beta+\beta-1) C^\top L^\top Q \\
        \ast &QA+A^\top Q-QLC-C^\top L^\top Q
    \end{smallmatrix}\right]\xi(t).
\end{equation}
For $\dot{V}$ to be negative definite,
\begin{equation}
      \begin{bmatrix}
       PA+A^\top P  &(\alpha\beta+\beta-1) C^\top L^\top Q \\
        \ast &QA+A^\top Q-QLC-C^\top L^\top Q
    \end{bmatrix}\prec 0.
    \end{equation}
    
 Let $QL=W$, then the above inequality becomes,
\begin{equation}
\begin{bmatrix} PA+A^\top P &(\alpha\beta+\beta-1)C^\top W^\top\\ \ast &QA+A^\top Q-WC-C^\top W^\top \end{bmatrix}\prec 0.
\end{equation}

Based on the above analysis, the system \eqref{Closed-loop_sys} is asymptotically stable. For the positivity of the system \eqref{Closed-loop_sys}, $(A-LC)$ should be Metzler matrix. Using Lemma \ref{lemma_metzler}, we have
\begin{align}
      ( A-LC)+\lambda I&\geq 0, \nonumber\\
       QA-WC+\lambda QI&\geq 0.
\end{align}
The proof is completed.
\endproof
A positive minimum bound of IET is discussed in Theorem \ref{theorem_2} in order to exclude Zeno behavior.
\begin{theo} \label{theorem_2}
The system \eqref{Closed-loop_sys} excludes Zeno behavior if IET satisfies the following relation 
\begin{equation}
    t_{k+1}-t_k\geq \frac{\alpha}{(\alpha+1)\|A\|}>0.
\end{equation}
\end{theo}
\proof
    Let $m(t)=\frac{\|\epsilon(t)\|}{\|y(t)\|}$, then for all $t\in [t_k,t_{k+1})$ we have
\begin{align}
\dot{m}(t) & =\frac{\dv{\|\epsilon(t)\|}{t}\|y(t)\|-\|\epsilon(t)\|\dv{\|{y}(t)\|}{t}}{\|y(t)\|^2}  \nonumber \\
\dot{m}(t)& \leq \frac{\dv{\|{y}(t)\|}{t}}{\|y(t)\|}+\frac{\|\epsilon(t)\|\dv{\|{y}(t)\|}{t}}{\|y(t)\|^2} \nonumber \\
\dot{m}(t)& \leq \frac{\dv{\|{y}(t)\|}{t}}{\|y(t)\|}+\frac{(\alpha \beta+\beta-1)\|y(t)\|\dv{\|{y}(t)\|}{t}}{\|y(t)\|^2} \nonumber \\
\dot{m}(t)& \leq \beta(\alpha +1) \frac{\dv{\|{y}(t)\|}{t}}{\|y(t)\|} \label{eq_20} .
\end{align}
Using \eqref{PS_eq}, we have the following 
\begin{align}
\frac{\dv{\|{y}(t)\|}{t}}{\|y(t)\|} & \leq \frac{\|A x(t)\|}{\|x(t)\|} \nonumber \\
\frac{\dv{\|{y}(t)\|}{t}}{\|y(t)\|}& \leq\|A\|.\label{eq_21}
\end{align}
Combining \eqref{eq_20} and \eqref{eq_21} yields,
\begin{equation} \label{dot_m_eq}
    \dot{m}(t) \leq \beta(\alpha+1)\|A\|.
\end{equation}
We have $\epsilon(t)=\beta y(t_k)-y(t)\geq 0$, thus
\begin{align}
\epsilon\left(t_k\right)&=\beta y\left(t_k\right)-y\left(t_k\right), \nonumber \\ \epsilon\left(t_k\right)&=(\beta-1) y\left(t_k\right).
\end{align}
It is obtained that
\begin{align}
m\left(t_k\right)=\frac{\left\|\epsilon\left(t_k\right)\right\|}{\left\|y\left(t_k\right)\right\|}=\beta-1 .
\end{align}
Integrating both sides of inequality \eqref{dot_m_eq}, one has
\begin{equation} \nonumber
\int_{t_k}^t\dot{m}(t) \mathrm{d} \tau \leq \int_{t_k}^t \beta(\alpha+1)\|A\| \mathrm{d} \tau.
\end{equation}
Thus, the upper bound of $m(t)$ can be computed as
\begin{align} \nonumber
    &m(t)-m\left(t_k\right)\leq ((\alpha\beta+\beta)\|A\|)\left(t-t_k\right),\\
    &m(t)\leq ((\alpha\beta+\beta)\|A\|)\left(t-t_k\right)+\beta-1 \nonumber.
\end{align}
At the next sampling instance $t=t_{k+1}$, the maximum bound can be calculated using the above relation,
\begin{equation} \label{e_1_eq}
    m(t) \leq[((\alpha\beta+\beta)\|A\|)\left(t_{k+1}-t_k\right)+\beta-1].
\end{equation}
From the event rule \eqref{eq:event_law}, it can be assured that at the next sampling instance, we have
\begin{equation} \label{e_2_eq}
   m(t) =\alpha \beta+\beta-1 .
\end{equation}
Substituting  $m(t)$ from the above equation in \eqref{e_1_eq},
we have

\begin{equation} \label{zeno_condition}
  \alpha \beta+\beta-1 \leq  ((\alpha\beta+\beta)\|A\|)\left(t_{k+1}-t_k\right)+\beta-1 .
\end{equation}
It follows from \eqref{zeno_condition} that
\begin{equation} \label{zeno_condition_final}
    t_{k+1}-t_k \geq \frac{\alpha }{(\alpha+1)\|A\|} .
\end{equation}

By design, $\alpha>0$ and $(\alpha+1)\|A\|>0$, which gives
\begin{equation}
    \frac{\alpha}{(\alpha+1)\|A\|}>0,
\end{equation}
thereby from \eqref{zeno_condition_final} one can get
\begin{equation} 
    t_{k+1}-t_k \geq\frac{\alpha}{(\alpha+1)\|A\|}>0 .
\end{equation}
Thus, the Zeno behavior is excluded. The proof is completed.
\endproof
\begin{remark}
    Theorem \ref{theorem_2} provides an insight of the connection between the minimum positive value of IET and the parameter $\alpha$. Let $\delta$ represent the minimum positive bound of IET. Based on Theorem \ref{theorem_2}, we have the ability to establish a function, denoted as $\mathcal{F}:\mathbb{R}_+\longrightarrow \mathbb{R}_+$, that maps the parameter $\alpha$ to $\delta$ as: $\mathcal{F}(\alpha)=\frac{\alpha}{(\alpha+1)\|A\|}
     =\frac{1}{(1+\frac{1}{\alpha})\|A\|}$.
   It can be observed that the function $\mathcal{F}(\alpha)$ decreases as $\alpha$ decreases within the range of $0<\alpha<1$, leading to more number of times the system updates, thereby making the conditions of Theorem \ref{theorem_1} more conservative in terms of communication efficiency.
\end{remark}

We have proposed an event-based positive state observer for positive systems. These findings have the potential to be applied in other disciplines. For example, the population changes of specific species, like pests, can be represented using positive systems \cite{WangDong2016}. The amount of juvenile pests can be approximated using a designed state observer, based on measurements of both immature and adult bugs. Another utilization can be observed in the field of clinical pharmacology. An example of this is that the way anaesthetic is distributed and transferred in the human body may also be explained using positive systems \cite{haddad2006adaptive}. It is generally recognized that the anaesthetic will be distributed throughout various parts of the human body, such as the fat, muscles, vessels, and organs.
Based on mass balance, the dynamics of anaesthetic dispersion can be modelled as a compartmental system, a specific form of positive system. A positive state observer that has been specifically designed can be employed to approximate the quantity of the substance in a specific compartment. Furthermore, our findings can also be extended to domains such as ecological dynamics \cite{farina2000} and epidemic models \cite{KHANAFER2016126}, among others. In the current work we employed our proposed event-based positive observer on a multitank water system (in Example \ref{ex_2}) for estimating the water levels in the three storage tanks.

\begin{remark}
 This article proposes an event-based positive observer for a linear positive autonomous system \eqref{PS_eq} to estimate the non-negative states. However, the proposed design approach can be extended to a linear positive system with inputs.   
\end{remark}

\section{Illustrative Example}
In this section, a numerical analysis is performed using simulation to validate the effectiveness of the proposed event-based positive observer design. To assess the performance of the proposed positive observer, we consider two simulation scenarios: (i) an academic example (Example 1), and (ii) a three-tank water system model (Example 2).
\begin{ex} \label{ex_1}
Consider the following autonomous positive linear system,
\begin{equation}
\begin{aligned}
\dot{x}(t)&=\begin{bmatrix}
   -1  &3\\0 &-1
\end{bmatrix}x(t), \\
y(t)&=\begin{bmatrix}
     1 &0 
\end{bmatrix}x(t).
\end{aligned}
\end{equation}
\end{ex}
For $\alpha=0.3$ and $\beta= 1.5$, using sufficient conditions in Theorem \ref{theorem_1} and solving them in MATLAB using the YALMIP toolbox, the feasible solutions $P$, $Q$, and observer gain matrix, $L$ obtained are: 
\begin{align*}
P&=\begin{bmatrix}
         0.3655  &0\\
         0    &1.1736
    \end{bmatrix}, \quad Q=\begin{bmatrix}
        0.4056   &0\\
         0    &0.9079
\end{bmatrix}\\
L&=\begin{bmatrix}
        0.9037 & 0
    \end{bmatrix}^\top.
    \end{align*}
\begin{figure}[ht!]
\centering
\subfigure[]{
\includegraphics[width=0.3\textwidth]{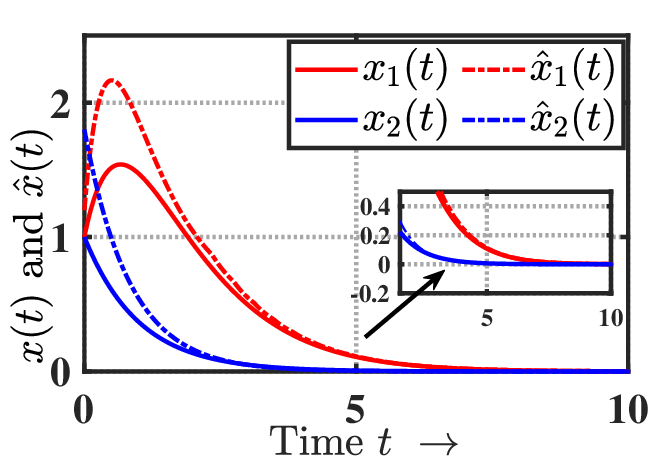}
\label{state_beta_0.3}}
\subfigure[]{
\includegraphics[width=0.3\textwidth]{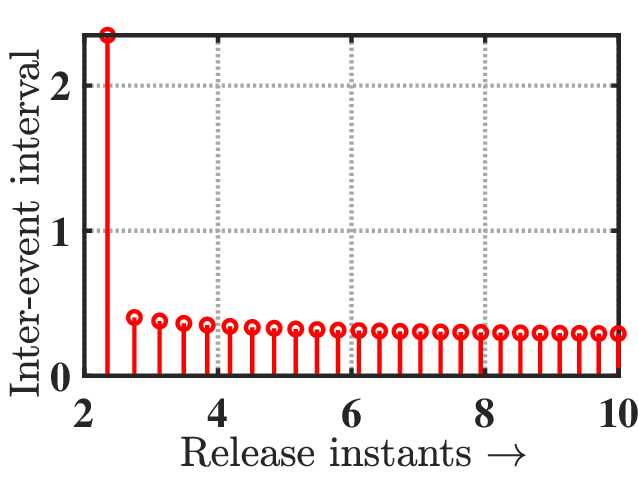}
\label{trigger_instants_beta_0.3}}
\caption{(a) States estimation and (b) Inter execution plot with $\alpha=0.3$ }
\end{figure}

The estimation of plant states, $x_1(t)$ and $x_2(t)$ with initial condition $x_0=[1.2 ~1.8]^\top$ are shown in Figure \ref{state_beta_0.3}. The plot for IET is shown in Figure \ref{trigger_instants_beta_0.3}. 
\begin{figure}[ht!]
\centering
\subfigure[]{
\includegraphics[width=0.3\textwidth]{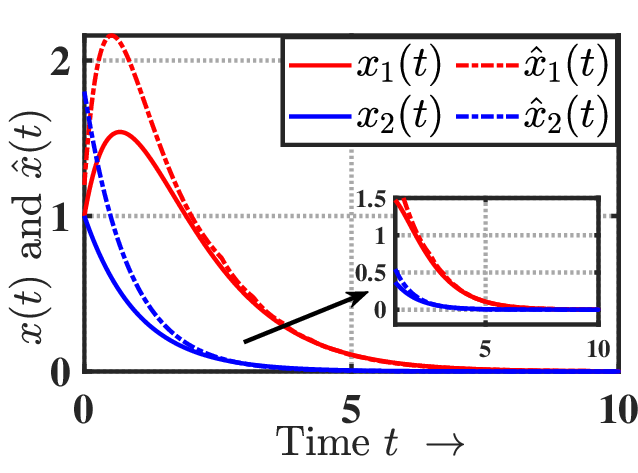}
\label{state_beta_0.5}}
\subfigure[]{
\includegraphics[width=0.3\textwidth]{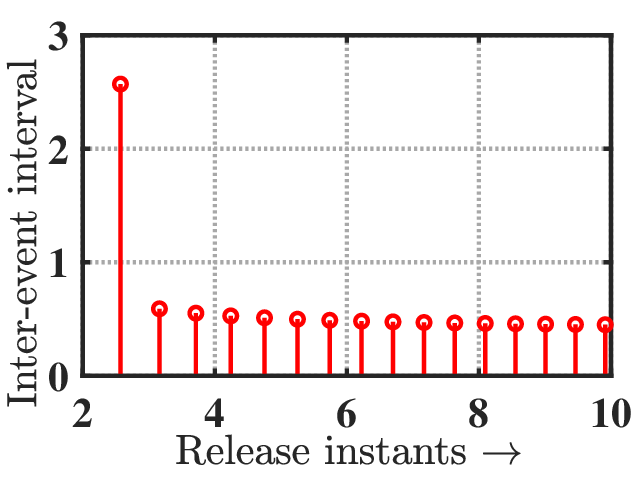}
\label{trigger_instants_beta_0.5}}
\caption{(a) States estimation and (b) Inter execution plot with $\alpha=0.5$}
\end{figure}

\begin{figure}[ht!]
\centering
\subfigure[]{
\includegraphics[width=0.28\textwidth]{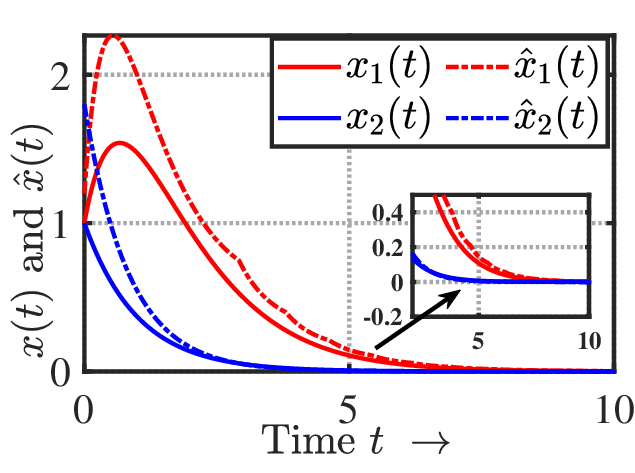}
\label{state_beta_0.9}}
\subfigure[]{
\includegraphics[width=0.28\textwidth]{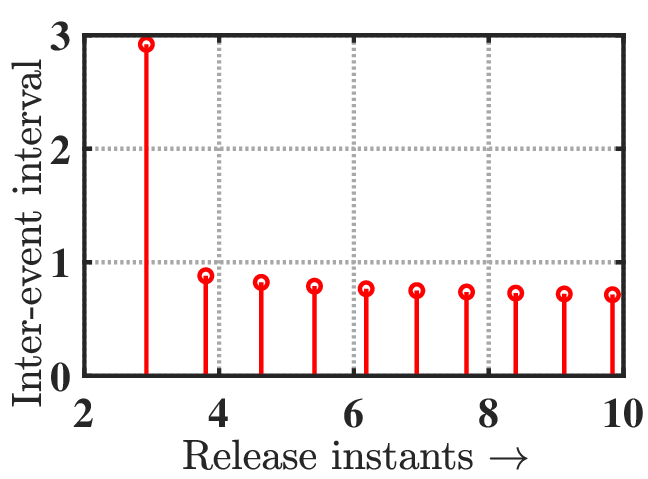}
\label{trigger_instants_beta_0.9}}

\caption{(a) States estimation and (b) Inter execution plot with $\alpha=0.9$}
\end{figure}

The impact of the change in the parameter $\alpha$ on the simulation can be observed in the following results. For different values of $\alpha=0.5,0.9$, the state estimation and IET are shown in Figure \ref{state_beta_0.5}-\ref{trigger_instants_beta_0.5} and Figure \ref{state_beta_0.9}-\ref{trigger_instants_beta_0.9} respectively. 
\begin{figure}[ht!]
\centering
\includegraphics[width=0.3\textwidth]{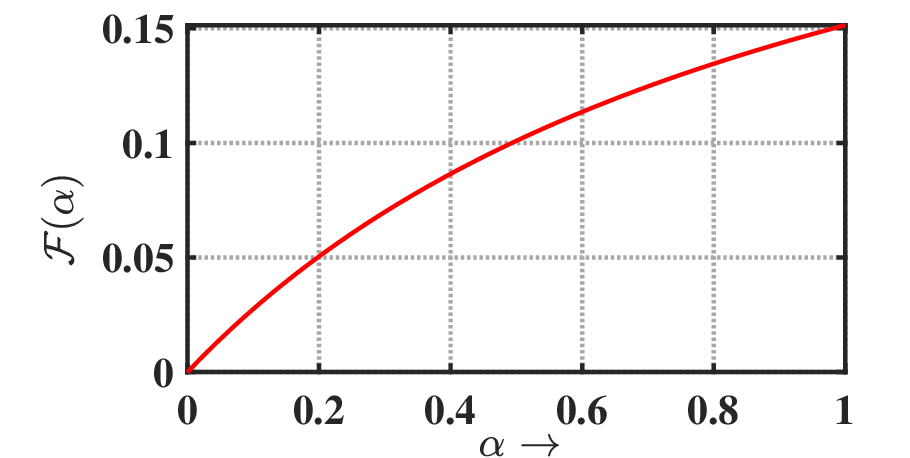}
\caption{Plot of minimum positive bound on IET, $\mathcal{F}(\alpha)$ vs event threshold, $\alpha$}
\label{EMT_alpha_vs_IET }
\end{figure}
It can be observed that with the increase in parameter $\alpha$, the communication efficiency increases, but system performance can be degraded. Thus, a trade-off should be there for choosing parameter $\alpha$ in practical applications. Also, the evolution of the positive lower bound on IET with respect to the change in threshold parameter $\alpha$ is shown in Figure \ref{EMT_alpha_vs_IET }. We can observe that with the increase in parameter $\alpha$, the positive bound also increases. Table \ref{table_1} gives a comparison of IET under different system parameters.

\begin{table}[h]
\small\sf\centering
\caption{Comparison of IET under different system parameters.\label{table_1}}
\begin{tabular}{llll}
\toprule
\textbf{$\alpha$}&\textbf{$\lambda$}&\textbf{Events} &Minimum positive \\&&&bound\\
\midrule
$0.3$ &$2.6341$ & $26$ & $0.0699$\\
$0.5$ &$2.8364$ & $17$ & $0.1009$\\
$0.9$ &$2.8833$ & $11$ & $0.1434$\\
$1$ &$2.5153$ & $10$ & $0.1514$\\
\bottomrule
\end{tabular}
\end{table}

\begin{figure}[ht!]
\centering
\includegraphics[width=0.5\textwidth]{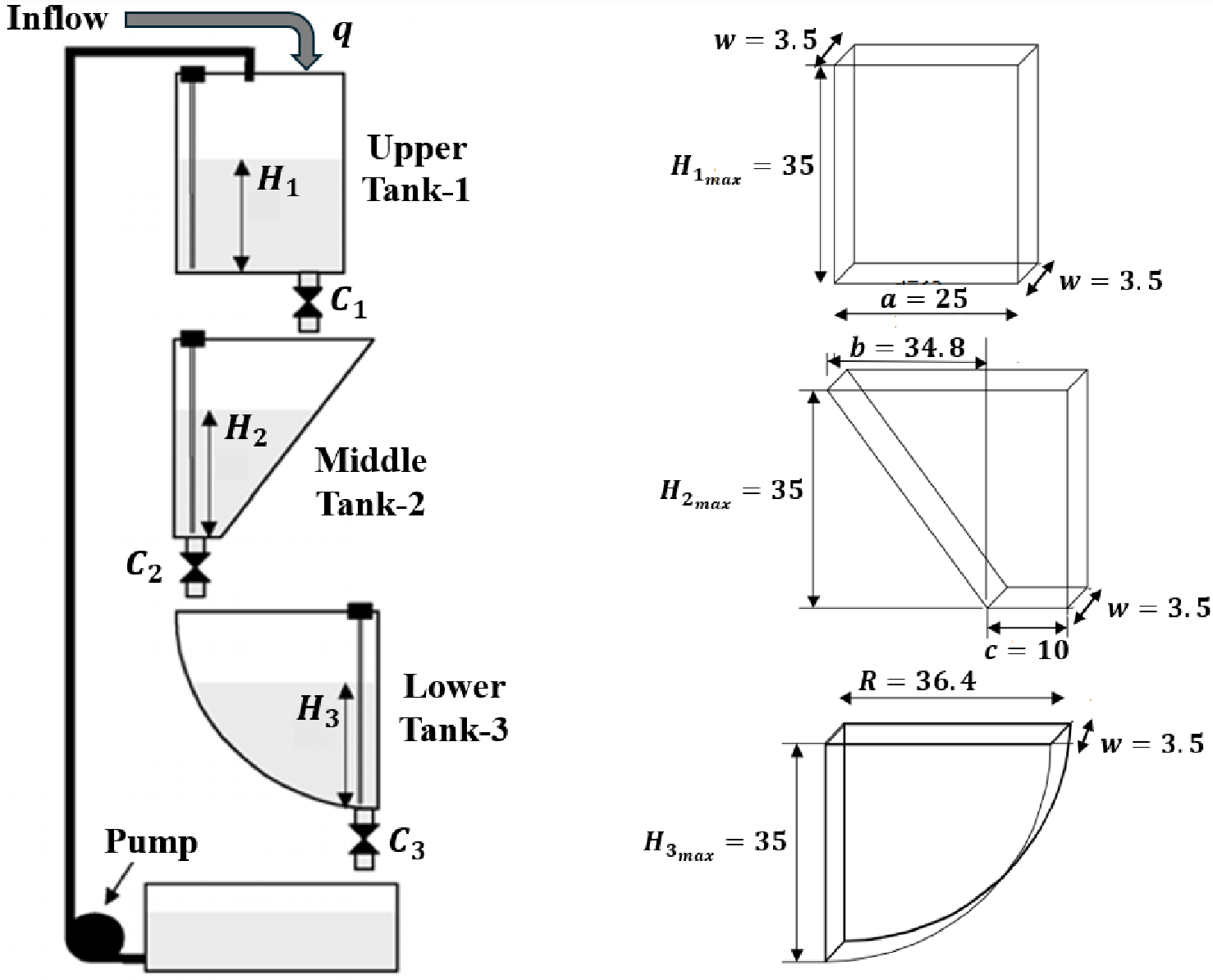}
\caption{Multitank configuration}
\label{multitank_sys}
\end{figure}

\begin{figure}[ht!]
\centering
\subfigure[]{
\includegraphics[width=0.50\textwidth]{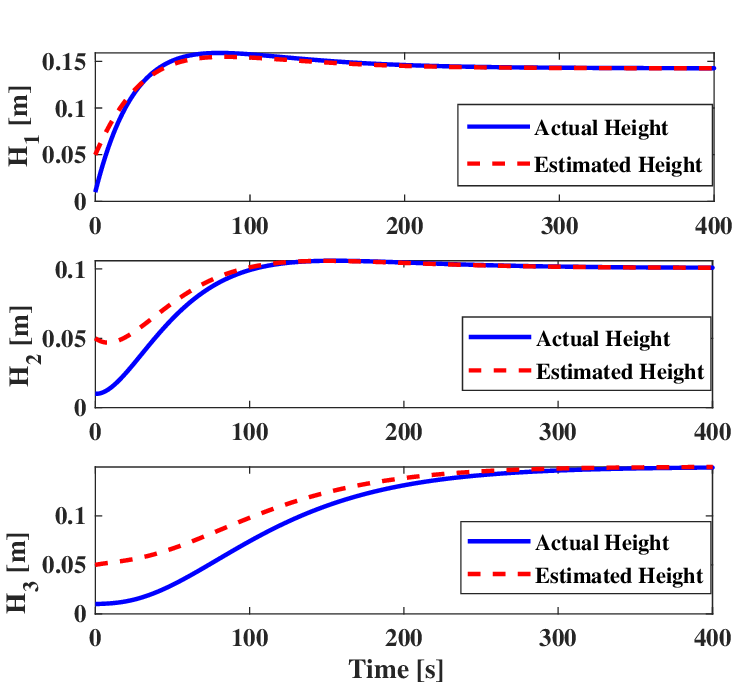}
\label{tank_sys_states}}
\subfigure[]{
\includegraphics[width=0.35\textwidth]{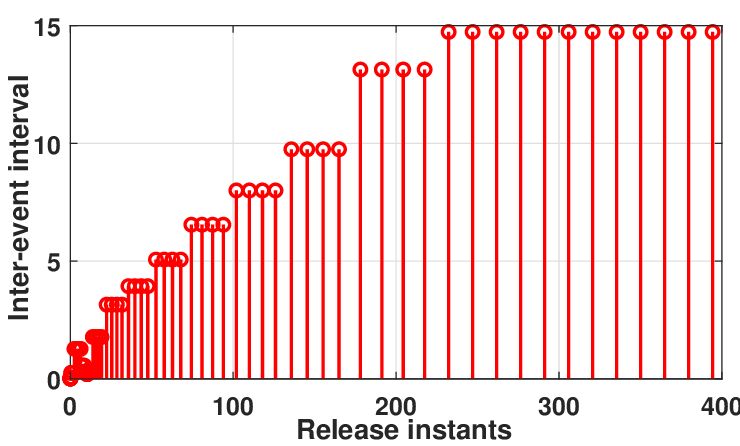}
\label{tank_IET_plot}}

\caption{(a) Water level estimates in all the three tanks and (b) Inter execution plot}
\end{figure}

\begin{ex} \label{ex_2}
The linearised dynamical model of the three tank system having variable cross-section \cite{inteco_multitank_manual} is described by:
\begin{equation}
\begin{aligned}
    \dot{h}(t)&=Ah(t)+Bu(t),\\ y(t)&=Ch(t)
    \end{aligned}
\end{equation}
where the matrices $A$ and $B$ are as follows:
\begin{equation*}
\begin{aligned}
A &=
\begin{bmatrix}
\frac{-C_1 \alpha_1}{\beta(H_1) H_1^{1 - \alpha_1}} & 0 & 0 \\
 \frac{C_1 \alpha_1}{\beta(H_2) H_1^{1 - \alpha_1}} & 
\frac{-C_2 \alpha_2}{\beta(H_2) H_2^{1 - \alpha_2}} & 0 \\
0 & 
 \frac{C_2 \alpha_2}{\beta(H_3) H_2^{1 - \alpha_2}} & 
 \frac{-C_3 \alpha_3}{\beta(H_3) H_3^{1 - \alpha_3}}
\end{bmatrix}_{H = H_0}\\
B&=
\begin{bmatrix}
\frac{1}{\beta(H_1)}\\ 0\\0 \end{bmatrix};\quad C=
\begin{bmatrix}
0 &1 &0 \end{bmatrix}
\end{aligned}
\end{equation*} with  $h=H-H_0$ where $H_0$ is the equilibrium state and 
$$
\begin{aligned}
H_i&-\text{fluid level in the $i$ tank}, i=1,2,3.\\
\alpha_i&- \text{flow coefficient of $i$ tank}\\
C_i&- \text{valve coefficient of $i$ tank}\\
\beta(H_1)&:\text{cross-sectional area of Upper tank-1}\\&=aw\\
\beta(H_2)&:\text{cross-sectional area of Middle tank-2}  \\&=w \left(c + b \frac{H_2}{H_{2_{\max}}}\right)\\\beta(H_3)&:\text{cross-sectional area of Lower tank-3}\\&=w \sqrt{R^2 - (H_{3_{\max}}-H_3)^2}
\end{aligned}
$$ 
The geometrical and simulation parameters of the tank model are given in Table \ref{table_2}. 

\begin{table}[h]
\small\sf\centering
\caption{System Parameters.\label{table_2}}
\begin{threeparttable}
\begin{tabular}{llll}
\toprule
\textbf{Parameter [cm]} & \textbf{Value} & \textbf{Parameter} & \textbf{Value}\\
\midrule
$a$  & 25 & $C_1$ & $1.0057e-004$ \\
$b$  & 34.8 & $C_2$ & $1.1963 e-004$ \\
$c$  & 10 & $C_3$ & $9.8008e-005$ \\
$w$  & 3.5 & $\alpha_i$ & $0.5$ \\
$R$  & 36.4 & $H_{1,0}$  & $0.1425$ m \\
$H_{2_{\max}}$  & 35 & $H_{2,0}$ & $0.1007$ m \\
$H_{3_{\max}}$ & 35 & $H_{3,0}$ & $0.1500$ m \\
\bottomrule
\end{tabular}
\vspace{0.3em}
\begin{tablenotes}
\footnotesize
\item[] cm = centimeters; m = meters
\end{tablenotes}
\end{threeparttable}
\end{table}

The control input is taken as: $$u=-\begin{bmatrix}
0.1983e-003 &0.0765e-003 &0.0496e-003\end{bmatrix}h.$$ 
\end{ex}
The steady-state inflow is $
    Q_0 = 3.7958e{-005} \, \text{m}^3/\text{s}$ and the initial conditions are $
    H_1(0) = H_2(0) = H_3(0) = 0.01 \, \text{m}$.
Taking $\alpha=0.5$ and $\beta= 1.5$, using sufficient conditions in Theorem \ref{theorem_1} and solving them in MATLAB using the YALMIP toolbox, the observer gain matrix, $L=\begin{bmatrix}
     0.0176 &0.0104 &0
  \end{bmatrix}^\top$. The estimation of the water level, $H_1$, $H_2$ and $H_3$ with initial condition $\hat{H}_0=[0.05 ~0.05 ~0.05]^\top$ are shown in Figure \ref{tank_sys_states}. 

 Figure \ref {tank_IET_plot} shows the transmission intervals, where only $85$ data packets are sent from the sensor to the network over the time period 
$t\in [0,400]s$. Thus, it is clearly observed that the desired water levels  $H_{1,0}, H_{2,0}, H_{3,0}$ are estimated by the proposed event-based positive observer with a substantial reduction in the transmission of the data packets as shown in Table \ref{tab:observer_comparison} compared to the traditional observer design with a periodic sampling scheme (e.g., at $1 s$ intervals requiring $400$ transmissions). MATLAB simulation reveals that this event-based design helps in a resource savings of approximately $78.75\%$, which highlights the efficacy and practicality of the proposed design compared to existing conventional design approaches.

\begin{table}[h]
\small\sf\centering
\caption{Comparison of System performance.\label{tab:observer_comparison}}
\begin{tabular}{ll}
\toprule
Implementation techniques&Total Sample\\& Updates\\
\midrule
Event based implementation&$85$\\
Traditional Periodic implementation &$400$\\
\bottomrule
\end{tabular}
\end{table}


\section{CONCLUSIONS}
A new positive state observer design for continuous-time linear positive networked systems is proposed in this work. A weight factor-based sampling error is used to construct a positive event-based measurement law. In order to ensure the positivity and asymptotic stability of the observed system, sufficient conditions are derived in terms of LMIs. Also, the event-based design is free from Zeno behavior, since a positive minimum bound is ensured on IET. The effectiveness of the proposed observer is validated through numerical simulations on a three-tank system having variable cross-section with significant savings in communication resources. The proposed observer can be useful for various applications, e.g., output feedback stabilization, fault detection, and isolation. However, in various practical scenarios, disturbances and uncertainties are inevitable, which can be treated as a future extension to the current work.





\section{Conflicts Of Interest}
The authors declare that there is no known competing financial interest or personal relationship that might have potentially impacted the work outlined in this paper.

\section{Funding}
No funding has been received to assist with this research work.





\bibliographystyle{SageH}

\bibliography{reference.bib}

\end{document}